\def\simless{\mathbin{\lower 3pt\hbox
             {$\rlap{\raise 5pt\hbox{$\char'074$}}\mathchar"7218$}}}    %~<

\def\simmore{\mathbin{\lower 3pt\hbox
             {$\rlap{\raise 5pt\hbox{$\char'076$}}\mathchar"7218$}}}    %~>

\documentclass[structabstract]{aa}

\usepackage{graphicx}
\begin{document}
\title{
The energy distribution of electrons in radio jets
}

\subtitle{}

\author{
Alexandros Tsouros\inst{1}
\and
Nikolaos D. Kylafis\inst{1,2}
}

\institute{
University of Crete, Physics Department \& Institute of
Theoretical \& Computational Physics, 71003 Heraklion, Crete, Greece\\
\and
Foundation for Research and Technology-Hellas, IESL,
71110 Heraklion, Crete, Greece\\
}

\date {Received ; Accepted ;}

%\abstract{}{}{}{}{}
% 5 {} token are mandatory

\abstract
% context heading (optional)
{
Black-hole and neutron-star X-ray binaries exhibit compact radio jets,
when they are in the so called quiescent, hard, or hard intermediate states.  
The radio spectrum in these states is flat to slightly inverted, i.e., 
the spectral index of the observed flux density is in the range
$0 \simless \alpha \simless 0.5$.  It is widely accepted that 
the energy distribution of the electrons, in the rest frame of the jet,
is a power law with index in the range $3 \simless p \simless 5$.
}
% aims heading (mandatory)
{
Contrary to what our thinking was decades ago,
now we know that the jets originate in the hot, inner 
flow around black holes and neutron stars.  So it is worth investigating 
the radio spectrum that is emitted by a thermal jet as a function
of direction.
}
% methods heading (mandatory)
{
As an example,
we consider a parabolic jet and, with the assumption of flux freezing,
we compute the emitted spectrum in all directions,
from radio to near infrared, using either a thermal distribution of electrons
or a power-law one.  
}
% results heading (mandatory)
{
We have found that parabolic jets with a thermal distribution of electrons give 
also flat to slightly inverted spectra.  
In particular, for directions along the jet ($\theta=0$),
both distributions of electron energies give $\alpha = 0 \pm 0.01$.  
The index $\alpha$ increases as the viewing angle $\theta$ increases and for
directions perpendicular to the jet ($\theta = \pi/2$), 
the thermal distribution gives
$\alpha = 0.40 \pm 0.05$, while the power-law distribution gives 
$\alpha = 0.20 \pm 0.05$. 
The break frequency $\nu_b$, which marks the transition from partially 
optically thick to optically thin synchrotron emission, is comparable
for the power-law and the thermal distributions.
}
% conclusions heading (optional)
{
Contrary to common belief, it is not necessary to invoke a power-law
energy distribution of the electrons in a jet to explain its flat to
slightly inverted radio spectrum.  A relativistic Maxwellian produces
similar radio spectra.  Thus, the jet may be the widely invoked ``corona''
around black holes in X-ray binaries.
}


\authorrunning{Tsouros \& Kylafis 2017}
\titlerunning{Energy distribution of electrons in radio jets}

\maketitle

%________________________________________________________________

\section{Introduction}

Black-hole X-ray binaries (BHXB) always exhibit a compact radio jet when 
they are in one of three spectral states:  
quiescent, hard, and hard intermediate
(Fender et al. 2004; Fender et al. 2009;
Fender \& Gallo 2014; Gallo et al. 2014). 
For a classification of the spectral states of BHXB see Belloni et al. (2005). 

For the formation of the jet, two mechanisms have been proposed:
plasma gun/magnetic tower (Contopoulos 1995; Lynden-Bell 1996)
and centrifugal driving (Blandford \& Payne 1982).  Both of 
them require a strong, large-scale, poloidal magnetic field.  Such a magnetic
field can either originate from a large distance 
from the black hole and the advecting flow 
carries it to the inner region and amplifies it (Igoumenshchev 2008;
Lovelace et al. 2009; Tchekhovskoy et al. 2011), or it can be produced locally 
by the Cosmic Battery (Contopoulos \& Kazanas 1998; see also
Contopoulos et al. 2006, Christodoulou et al. 2008, 
Contopoulos et al. 2009; Contopoulos et al. 2015).
We favor the Cosmic Battery, because jets 
in BHXB are destroyed and re-created within
hours, when the sources cross the so-called jet line (Fender et al. 2004;
Miller-Jones et al. 2012).  
We consider it highly unlikely that the sources anticipate the destruction 
of their jet and its subsequent re-formation so as to ``request'' 
a magnetic field from far away, which should arrive at the inner part of the 
flow at the time that it is needed.  Instead,
we think that the observations require the strong poloidal
magnetic field to be produced locally, at the right place and the right time.
The formation and destruction of jets
in the context of the Cosmic Battery, as well as the relevant
timescales, have been discussed in Kylafis et al. (2012).  An explanation of 
the rich phenomenology during an outburst of a BHXB has been offered by
Kylafis \& Belloni (2015a; 2015b).

The spectra of BHXB from the radio to the near infrared  
are flat to slightly inverted, i.e. flux density
$S_{\nu} \propto \nu^{\alpha}$, with $0 \simless \alpha \simless 0.5$
(Mirabel \& Rodriguez 1999;
Fender et al. 2000; 
Fender 2001;
Fender et al. 2001;
Russell et al. 2006;
Corbel et al. 2013;
Russell \& Shahbaz 2014;
Tetarenko et al. 2015).
A characteristic frequency in this spectrum is the break frequency $\nu_b$,
where the partially optically thick jet becomes optically thin.  This 
is an important frequency, because $S_{\nu_b}$ is an indication of the total
power emitted by the jet.  The frequency $\nu_b$ varies from source to source
(Russell et al. 2013a) and also for the same source as a function of time
(Russell et al. 2013b; Russell et al. 2014).
Above $\nu_b$, the jet is optically thin and its spectrum falls with an index
$-1 \le \alpha \le -0.5$ (Fender 2001).

In 1979, synchrotron radio spectra were calculated for both, a Maxwellian
distribution of electrons (Jones \& Hardee 1979) and for a power-law one
Blandford \& K\"{o}nigl (1979).  In subsequent years, the power-law model
became the standard one 
and the research efforts concentrated on explaining how a power-law 
distribution of electron energies can be produced.

Two mechanisms can produce power-law distributions of electron energies:
shocks (Heavens \& Meisenheimer 1987; for a review see Drury 1983) 
and magnetic reconnection 
(Spruit et al. 2001; Drenkhahn \& Spruit 2002;
Sironi \& Spitkovsky 2014;
Sironi et al. 2015;
for a review see Kagan et al. 2015).  
The question
then arises:  are shocks and/or magnetic reconnection guarranteed to be
present in the {\it entire} jet, from its base (where the frequency $\nu_b$
is determined) to the top?  
One can envision shocks, due to an uneven flow in the jet,
and magnetic reconnection, due to partially turbulent 
magnetic fields, but it is hard to imagine that these mechanisms operate
in the entire jet.

In recent years, it has been accepted 
(Fender 2006; Fender \& Gallo 2014; 
Kylafis et al. 2012; Kylafis \& Belloni 2015a, b)
that the jets in BHXB originate in the geometrically thick, 
optically thin, hot, inner flow around black holes, where the temperature
of the electrons is large (hundreds of keV)
(Ichimaru 1977; Narayan \& Yi 1994; 1995; Abramowicz et al. 1995; 
Blandford \& Begelman 1999; Narayan et al. 2000; Qataert \& Gruzinov 2000;
Yuan et al. 2005).  
Outside the hot flow, the accretion disk is
radiatively efficient and geometrically thin, i.e. Shakura-Sunyaev - type. 
The transition radius between the two types of flow decreases with 
increasing mass
accretion rate and the Shakura-Sunyaev disk extends all the way to the
inner stable circular orbit at high accretion rates, when the sources
are in the so called soft state and no jet is present 
(Fender, et al. 1999; Russell et al. 2011).

In this picture, it is natural to expect that the electrons in the jet,
at least at its base, should be thermal or close to thermal.
Therefore, even out of curiosity, we examine what type of radio spectrum
is produced by a thermal distribution of electrons in the jet.  

As mentioned above, most of the theoretical work on the radio emission 
from jets has assumed a power-law distribution of electron energies
(see however Falcke \& Markoff 2000, who considered also a thermal
jet model for Sgr A*).
An extensive study of jet radio spectra using a power-law distribution 
was done by Kaiser (2006).  An also extensive study, using both power-law 
and thermal distributions, was done by Pe'er \& Casella (2009).  In order 
to obtain analytic results, both of these studies calculated spectra 
in the direction perpendicular to the jet ($\theta=\pi/2$).  
However, as we show below, 
the spectral index $\alpha$ depends strongly on the observation 
angle $\theta$.

In this {\it Letter} we consider a simple jet model, compute the radio 
spectrum as a function of $\theta$, 
and demonstrate that two completely different 
electron energy distributions (thermal and power-law)
result in similar spectra.  In subsequent work,
we will explore different models to see if it is possible to infer the
electron energy distribution from observations.

In \S~2 we describe our model,
in \S~3 we compute the radio emission from the jet,
in \S~4 we remark on some aspects of our calculations, 
and in \S~5 we present our conclusions.

\section{The model}

\subsection{Characteristics of the jet}

As a demonstration, we assume a parabolic jet, which has been used
extensively before (see \S~4).  
In other words, 
we assume that the radius of the jet as a function of distance from the
center of the compact object is
$$
R(z)= R_0 ~ ( z/z_0)^{1/2},
\eqno(1)  
$$
where $R_0$ is the radius at the base of the jet and $z_0$ is the height 
of the base of the jet.  

For simplicity, we assume that the jet is accelerated close to its launching 
region and that it has constant velocity $v_{\parallel}=0.8 c$.  
From the continuity equation we infer that the number density of the 
electrons in the jet as a function of distance is
$$
n_e(z)= n_0 ~ z_0/z,
\eqno(2)
$$
where $n_0$ is the number density of the electrons at the base of the jet.

For the magnetic field in the jet, we also make the simple assumption
that it is nearly parallel to the $z-$axis and that its strength
is determined by flux conservation along the jet: $B(z) \pi R^2(z) =$ 
constant.  This implies that
$$
B(z)=B_0 ~ z_0/z,
\eqno(3)
$$
where $B_0$ is the strength of the magnetic field at the base of the jet.

\subsection{Electron energy distribution}

For the energy distribution of the electrons in the rest frame of the 
jet, or equivalently for the distribution of the Lorentz factor 
$\gamma$ since $E_e = \gamma m_e c^2$, we assume either a relativistic
Maxwellian, i.e. a Maxwell-J\"{u}ttner distribution,
$$
f_{\rm MJ}(\gamma)={ {\gamma^2 \beta} \over {\Theta K_2(1/\Theta)} }
e^{-\gamma/\Theta},
\eqno(4a)
$$
where $\beta= \sqrt{1-1/\gamma^2}$, $\Theta =kT_e/m_ec^2$, and
$K_2$ is the modified Bessel function of the second kind,
or a power-law distribution
$$
f_{\rm pl}(\gamma)={ {p-1} \over {\gamma_{\rm min}^{-p+1} -
\gamma_{\rm max}^{-p+1}} } \gamma^{-p}, 
\eqno(4b)
$$
in the range 
$\gamma_{\rm min} \le \gamma \le \gamma_{\rm max}$.

An analytic treatment of cooling in a jet was 
described in Kaiser (2006). For the values of the parameters that we use, 
the synchrotron timescale at, say $z=10z_0$, is about 1 s, while
the flow timescale there is 0.003 s.  Thus, synchrotron cooling can be
neglected.  For simplicity, we also neglect diabatic expansion cooling, 
because $\gamma(z) \propto (z_0/z)^{1/3}$ (Pe'er \& Casella 2009).

In Fig. 1 we plot the Maxwell-J\"{u}ttner distribution for $\Theta = 0.4$ 
as a line with stars and a power law distribution
with $p=4$ from $\gamma_{\rm min} = 1$ to
$\gamma_{\rm max} = 10$ as a line with diamonds.
Both distributions are normalized to unity.
Despite the fact that the distributions described by eqs. (4a) and (4b) 
are qualitatively different,
in the range $1 \simless \gamma \simless 6$ the two distributions
exhibit only quantitative differences.

\begin{figure}
\begin{center}
\includegraphics[angle=-90,width=8cm]{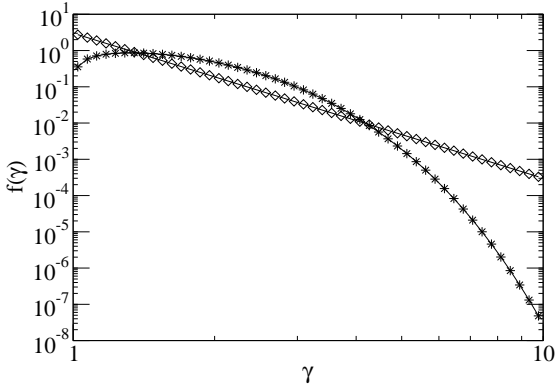}
\end{center}
\caption[]{
Plot of the Maxwell- J\"{u}ttner distribution (stars) for
$\Theta= kT_e/m_ec^2=0.4$ and of the power-law distribution (diamonds) for
$p=4$ and $\gamma_{\rm min} = 1$, $\gamma_{\rm max} = 10$.
}
\label{Fig1}
\end{figure}

The distribution of electrons as a function of $z$ and $\gamma$ is then
$$
n_e(z, \gamma)= n_e(z) f(\gamma),
\eqno(5)
$$
where $n_e(z)$ is given by eq. (2) and $f(\gamma)$ is given by eq. (4a) or
(4b).

\subsection{Radiative transfer}

Ignoring possible Compton upscattering in the jet
(for this the reader is referred to
Reig et al. 2003;
Giannios et al. 2004;
Giannios 2005;
Kylafis et al. 2008;
Reig \& Kylafis 2015; 2016
),
the equation for the transfer of radio photons in the jet 
in direction $\hat n$, along which length is measured by $s$, 
is given by
$$
{{dI(\nu, s)} \over {ds}} = j(\nu, s) - a(\nu, s)I(\nu,s),
\eqno(6)
$$
where $j(\nu,s)$ and $a(\nu, s)$ are the emission and absorption coefficients,
respectively.  

For mildly relativistic magnetic plasma, the gyrosychrotron 
emission formalism is the appropriate one.  
For simplicity, here we use the extreme relativistic formalism, in which
the emission coefficient is (see eq. 6.36 of Rybicki \& Lightman 1979)
$$
j(\nu, z)= n_e(z) D(z) \int_{\gamma_{\rm min}}^{\gamma_{\rm max}}
F\left( \nu \over \nu_c \right)
f(\gamma) d\gamma,
\eqno(7)
$$
where 
$$
D(z)= { {\sqrt{3} e^3 B(z) \sin \phi} \over {m_e c^2} }, 
$$
$e$ is the charge of the electron,
$\phi$ is the pitch angle of the electron, 
$$
\nu_c = { {3 \gamma^2 e B(z) \sin \phi} \over {4 \pi m_e c} },
$$
and
$$
F(x)= x \int_x^\infty K_{5 \over 3}(u) du,
$$
where $K_{5 \over 3}$ is the modified Bessel function of the second kind.

\begin{figure}
\begin{center}
\includegraphics[angle=-90,width=8cm]{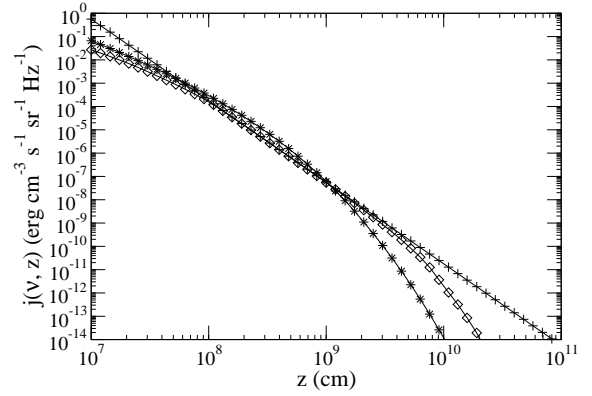}
\end{center}
\caption[]{
Emission coefficient $j(\nu, z)$ as a function of $z$ for 
$\nu = 10^{11}$ Hz.  The electron energy distributions are: thermal (stars), 
power-law with $1 \le \gamma \le \infty$ (plusses), and 
power-law with $1 \le \gamma \le 10$ (diamonds).
}
\label{Fig2}
\end{figure}

The absorption coefficient is (see eq. 6.50 of Rybicki \& Lightman 1979)
$$
a(\nu, z) = - { {n_e(z) D(z)} \over {8 \pi m_e \nu^2} }
\int_{\gamma_{\rm min}}^{\gamma_{\rm max}}
F\left( \nu \over \nu_c \right)
\gamma^2
{\partial \over {\partial \gamma} }
\left[{{f(\gamma)} \over {\gamma^2} } \right]
d\gamma.
\eqno(8)
$$

The formal solution to the radiative transfer equation (6), for a
distant observer, is
$$
I(\nu) = \int_{s_b}^{s_f} j(\nu, s) ds ~ ~
e^{-\int_{s}^{s_f} a(\nu, s^{\prime}) ds^{\prime}},
\eqno(9)
$$
where $s_b$ and $s_f$ are the back and the front intercepts of the line of 
sight with the surface of the jet.

The radio spectra of jets are observed to have a characteristic break 
frequency  $\nu_b$
(see Pe'er \& Casella 2009 for other possible characteristic frequencies), 
which is determined by the condition
that the jet has optical depth equal to one at its base
$$
a(\nu_b, z_0) R_0 = 1,
\eqno(10)
$$
where $R_0$ is the radius of the jet at its base (see eq. 1).  
This means that the entire jet is optically thin.  Due to contamination
of the radio spectrum, $\nu_b$ may not be 
always evident in the observed spectra.

\section{Radio emission from the jet}

For the purposes of a demonstrative calculation, we assign reference 
values to the involved parameters.  Thus, we take 
$R_0 = 100 r_g$, where $r_g=15$ km is the gravitational radius,
$z_0 = 5 r_g$,
$n_0 = 10^{16}$ cm$^{-3}$,
$B_0 = 2 \times 10^5$ G,
$\Theta = 0.4$,
$p = 4$, 
$\gamma_{\rm min} = 1$, and
$\gamma_{\rm max} = \infty$.  For the pitch angle we take the
value $\phi = 30$ degrees.

For the reference values of the parameters we find using eq. (10) that
$\nu_b = 2.7 \times 10^{14}$ Hz for the power-law distribution of 
electron energies and
$\nu_b = 10^{14}$ Hz for the thermal distribution.  
If we restrict the power-law 
distribution to $\gamma_{\rm max} = 10$, 
then $\nu_b = 1.6 \times 10^{14}$ Hz.

In Fig. 2 we show the emission coefficient $j(\nu, z)$ as a function of
$z$, for frequency $\nu = 10^{11}$ Hz.
The symbols are: stars (thermal) and plusses (power-law).
If we restrict the power-law 
distribution to $\gamma_{\rm max} = 10$, then the corresponding line 
becomes that with the diamonds.
It is evident from Fig. 2 that a power-law distribution with 
$\gamma_{\rm min} = 1$ and $\gamma_{\rm max} = 10$
has essentially the same emission coefficient as a thermal distribution. 

We have solved numerically expression (9) and, for the reference values of 
the parameters, we have found the following values for the spectral index:
$\alpha = 0$ {\it for both} electron energy distributions if the viewing
angle $\theta = 0 $.  For $\theta = \pi/2$, we have found 
$\alpha = 0.4$ for the thermal distribution and
$\alpha = 0.2$ for the power-law distribution, the last one
in agreement with Giannios (2005).  If we restrict
the power-law distribution to $1 \le \gamma \le 10$, then we find
$\alpha = 0.4$, something that was expected in view of the
similarity of the emission coefficient for the two distributions.
The value of $\alpha$ for both distributions increases smoothly from 
$\theta = 0$ to $\theta = \pi/2$. The variation of $\alpha$
with $\theta$ ($\Delta \alpha = 0.4$) 
is much larger than the variation 
($\Delta \alpha = \pm 0.05$)
caused by 
$0.2 \le \Theta \le 0.5$ or $3 \le p \le 5$. 
Also, there is no need to examine directions 
$\pi/2 < \theta \le \pi$, because the approaching lobe of the jet
dominates the emission.

\section{Discussion}

Models of BHXB invoke a thermal corona near the black hole to explain
the observed hard X-ray spectrum by thermal Comptonization(see Done
et al. 2007, for a review).  Here, we have demonstrated that a thermal
jet is an excellent ``corona'' for the upscattering of soft photons.
In fact, Comptonization in the jet can explain not only 
the observed power-law
high-energy spectral index $\Gamma$
and the high-energy cutoff $E_c$, but also
a) the depedence of the time lag on Fourier frequency (Reig et al. 2003),
b) the narrowing of the autocorrelation function with increasing photon
energy (Giannios et al. 2004), 
c) the dependence of $\Gamma$ on either the time lag or the
Lorentzian peak frequency (Kylafis et al. 2008), and 
d) the relation between $E_c$ and phase lag (Reig \& Kylafis 2015).
It is very interesting that all the above correlations are explained
with the same, simple, jet model, that has been used here as an example.
In fact, this was one of the reasons for choosing this example.

\section{Conclusions}

Contrary to common belief that the flat to slightly inverted radio spectra
from jets imply a power-law energy distribution of the electrons in the jet,
we have demostrated here that a thermal distribution of electron energies
produces essentially identical radio spectra.

We find it interesting that the index $\alpha$ of the radio spectrum of the
outbursting source MAXI J1836-194 flip-flops between $\sim 0.2$ and $\sim 0.5$
(Russell et al. 2014).  It is too early 
to infer that the energy distribution of the electrons alternates between
thermal and power law, but it is intriguing.

\begin{acknowledgements}
We thank an anonymous referee for useful comments and suggestions.
We also thank Dimitrios Giannios for useful discussions and comments.  One of
us (NDK) thanks Tomaso Belloni, Rob Fender, Sera Markoff, and Dave Russell
for useful discussions.

\end{acknowledgements}

\end{document}